\documentclass[aps,prd,twocolumn,groupedaddress,amsfonts,amssymb,amsmath,showpacs,showkeys]{revtex4-1}

\usepackage{mathtools}

\setcitestyle{authoryear,round}

\usepackage{graphicx}

\usepackage{psfrag}

\newcommand{\aap}{Astronomy and Astrophysics}

\newcommand{\mnras}{MNRAS}

\newcommand{\physrep}{Physics Reports}

\begin{document}

% THIS IS THE TITLE OF THE ARTICLE:
\title{Evidence for dark matter in the inner Milky Way...\textbf{Really?}}

\author{R. Durazo, X. Hernandez \& S. Mendoza}
\email[Email address: ]{rdurazo,xavier,sergio@astro.unam.mx}
\affiliation{Instituto de Astronom\'{\i}a, Universidad Nacional
                 Aut\'onoma de M\'exico, AP 70-264, Distrito Federal 04510,
	         M\'exico \\
            }

\date{\today}

% PACS numbers:
% \pacs{95.30.Sf,04.50.Kd,04.25.Nx,98.62.Dm,98.62.Sb}
\keywords{Galactic Dynamics;Relativity and gravitation;Modified theories of
gravity;Milky Way}

\begin{abstract}
  The following is a comment on the recent letter by \citet{iocco} where
the authors claim to have found ``\emph{...convincing proof of the
existence of dark matter...}''.  The letter in question presents a
compilation of recent rotation curve observations for the Milky Way,
together with Newtonian rotation curve estimates based on recent baryonic
matter distribution measurements.  A mismatch between the former and the
latter is then presented as ``\emph{evidence for dark matter}''.  Here we
show that the reported discrepancy is the well known gravitational anomaly
which consistently appears when dynamical accelerations approach the critical
Milgrom acceleration \( a_0 = 1.2 \times 10^{-10} \textrm{m} \,
\textrm{s}^{-2} \).  Further, using a simple modified gravity force law,
the baryonic models presented in \citet{iocco}, yield dynamics consistent
with the observed rotation values.
\end{abstract}

\maketitle

 The claim of ``Evidence for dark matter'' on a recent
letter to Nature Physics \citep{iocco} appears excessive.  The authors have
convincingly shown that the baryonic matter distribution in our galaxy
cannot account for the observed rotation curve of our galaxy, 
at scales somewhat shorter than those of the \( \sim 8 \textrm{kpc}
\) solar radius.  This result extends inwards the inconsistency
between the observed baryonic matter and the measured rotation curve,
already well known at large radii.  Two generic ways to deal with
this discrepancy are currently under discussion in the scientific
literature \citep{dm,benoit,odintsov,capozziello}: 
(a) To keep Newton's gravity unchanged
and make up any dynamical mismatch through the addition of as much
hypothetical non-baryonic dark matter as required.  And (b) To search
for a modified theory of gravity under which no such discrepancies
appear. The latter requires a transition away from Newton's gravitation
appearing below acceleration scales \citep{milgrom} \( a_0 \approx 1.2 \times
10^{-10}~\textrm{m}\, \textrm{s}^{-2} \).

 The figure shows how the discrepancies found by the authors appear
at an acceleration scale of order \( \ a_0 \).  Also shown in the
figure is the expected range of angular frequencies corresponding to
the baryonic distribution considered in the letter in question, under
a MOdified Newtonian Dynamics (MOND) model.  The discrepancy
is no longer evident, specially at galactocentric distances smaller than
\( 10 \textrm{kpc} \), acknowledged by the authors themselves as the high
quality data region.  Considering various of the MOND interpolation
functions proposed in the literature would somewhat broaden the cross
shaded region in the above plot, as it is precisely in this transition
region -where accelerations are of order \( a_{0} \)- that the various
possible interpolation functions differ.  We have also not included
uncertainties in the empirical calibrations \citep{gentile} of \( a_{0}
\) of \( \pm 0.3 \times 10 ^{-10 } \textrm{m}\, \textrm{s}^{-2} \),
since we aim merely to show how the velocity data presented by 
\citet{iocco} can be reproduced from the baryonic models these same
authors use.

The history of gravitational anomalies extends back to almost 200
years; when reporting on the observed residuals in the orbit of Neptune,
\citet{bouvard} correctly concluded that either (i) the effect of the
Sun's gravity, at such a great distance might differ from Newton's
description, or ( ii) the discrepancies might simply be observational
error; or ( iii) perhaps Uranus was being pulled, or perturbed, by
an as-yet undiscovered planet.  On that occasion option (iii) proved
correct, not so however in the following instance, where the observed
peculiarities in the orbit of Mercury turned out not to signal ``dark
matter'', but indeed, marked the end of the validity regime of Newtonian
gravity towards high velocities.

\begin{figure*}
  \includegraphics[scale=0.50]{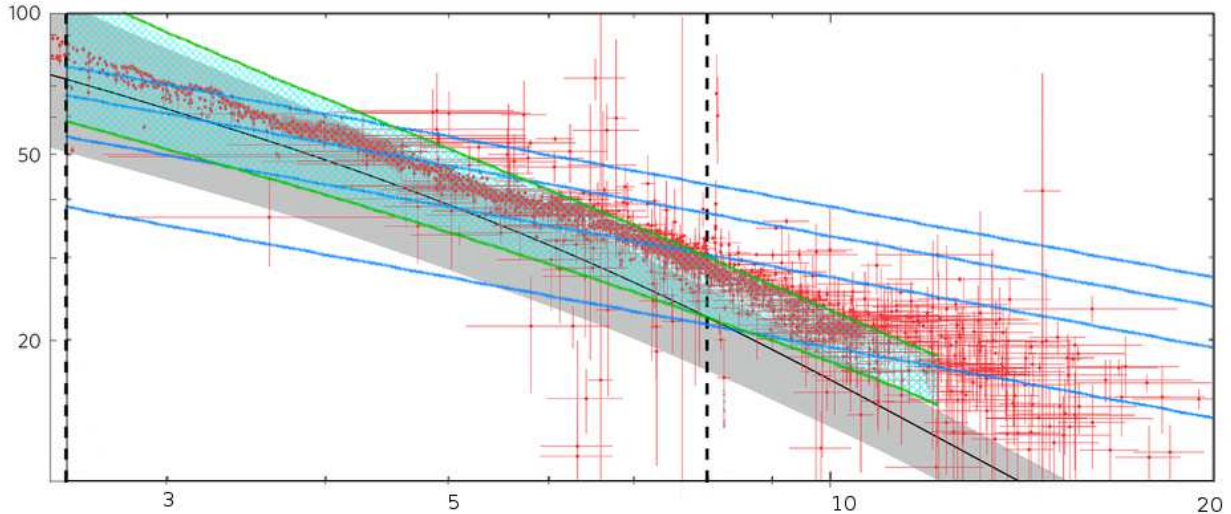}
  \caption[Figure]{ 
On the log-log angular frequency vs.
galactocentric radius plot of the upper panel from figure~(2) of 
\citet{iocco}, we have superimposed blue curves of constant
acceleration \( a = a_0,\ 2 a_0,\ 3 a_0,\ 4 a_0 \), from bottom
to top respectively.  The cross shaded region bound in green,
shows the angular frequencies which the gray baryonic models of the
above authors result in, using a MOND model \citep{mendoza} where the
gravitational force per unit mass in units of \( a_0 \) is given by \( f(x)
= \left( x^3 + x^2 + x \right)  / \left( x + 1 \right) \), where \( x^2 := G
M(R) / a_0 R^2 \) and \( G \) is Newton's gravitational constant, \( R \)
the galactocentric distance and \( M(R) \) the enclosed mass.  Note that
the Newtonian gravitational regime is recovered for \( x \gg 1 \).
}

\end{figure*}

 It is important to note that the difference between the two points of
view is far from merely semantic: both reflect fundamentally distinct
ideas of reality, either space is teeming with unseen particles far
outnumbering the detectable universe, or it is not. Both lead to distinct
predictions in a number of cases, e.g. black hole growth rates will be
affected by the accretion of dark matter, or not \citep{hernandez10}.
A satellite galaxy orbiting within a dark matter halo will gravitationally
interact with countless dark matter particles, loose energy and experience
dynamical friction, or it will not \citep{hernandez06}. Any theory
where the driving causal entity is something no one has ever seen,
(e.g. Cartesian vortices, phlogiston, caloric or the electromagnetic
aether) should be treated, at best, as a temporary working hypothesis.

In summary, Iocco's (2015) conclusion for  ``...a convincing
proof of the existence of dark matter...'' is misleading, specially
given that they fail to mention that their analysis is restricted to a
small subset amongst the many theories of gravitation currently under
consideration in the scientific literature.

%%%%%%%%%%%%%%%%%%%%%%%%%%%%%%%%%%%%% 

%%%%%%%%%%%%%%%%
% BIBLIOGRAPHY %
%%%%%%%%%%%%%%%% 
% \bibliographystyle{mn2e}
\bibliographystyle{aipauth4-1.bst}
\bibliography{dm-milky-really}

%merlin.mbs aipauth4-1.bst 2010-07-25 4.21a (PWD, AO, DPC) hacked
%Control: author (9) reversed initials
\begin{thebibliography}{11}%
\makeatletter
\providecommand \@ifxundefined [1]{%
 \@ifx{#1\undefined}
}%
\providecommand \@ifnum [1]{%
 \ifnum #1\expandafter \@firstoftwo
 \else \expandafter \@secondoftwo
 \fi
}%
\providecommand \@ifx [1]{%
 \ifx #1\expandafter \@firstoftwo
 \else \expandafter \@secondoftwo
 \fi
}%
\providecommand \natexlab [1]{#1}%
\providecommand \enquote  [1]{``#1''}%
\providecommand \bibnamefont  [1]{#1}%
\providecommand \bibfnamefont [1]{#1}%
\providecommand \citenamefont [1]{#1}%
\providecommand \href@noop [0]{\@secondoftwo}%
\providecommand \href [0]{\begingroup \@sanitize@url \@href}%
\providecommand \@href[1]{\@@startlink{#1}\@@href}%
\providecommand \@@href[1]{\endgroup#1\@@endlink}%
\providecommand \@sanitize@url [0]{\catcode `\\12\catcode `\$12\catcode
  `\&12\catcode `\#12\catcode `\^12\catcode `\_12\catcode `\%12\relax}%
\providecommand \@@startlink[1]{}%
\providecommand \@@endlink[0]{}%
\providecommand \url  [0]{\begingroup\@sanitize@url \@url }%
\providecommand \@url [1]{\endgroup\@href {#1}{\urlprefix }}%
\providecommand \urlprefix  [0]{URL }%
\providecommand \Eprint [0]{\href }%
\providecommand \doibase [0]{http://dx.doi.org/}%
\providecommand \selectlanguage [0]{\@gobble}%
\providecommand \bibinfo  [0]{\@secondoftwo}%
\providecommand \bibfield  [0]{\@secondoftwo}%
\providecommand \translation [1]{[#1]}%
\providecommand \BibitemOpen [0]{}%
\providecommand \bibitemStop [0]{}%
\providecommand \bibitemNoStop [0]{.\EOS\space}%
\providecommand \EOS [0]{\spacefactor3000\relax}%
\providecommand \BibitemShut  [1]{\csname bibitem#1\endcsname}%
\let\auto@bib@innerbib\@empty
%</preamble>
\bibitem [{\citenamefont {Bouvard}(1821)}]{bouvard}%
  \BibitemOpen
  \bibfield  {author} {\bibinfo {author} {\bibnamefont {Bouvard}, \bibfnamefont
  {A.}},\ }\href {http://books.google.com.mx/books?id=LPRVAAAAcAAJ} {\emph
  {\bibinfo {title} {Tables Astronomiques Publiees Par Le Bureau Des Longitudes
  De France, Contenant Les Tables De Jupiter, De Saturne Et D'Uranus,
  Construites D'Apres La Theorie De La Mecanique Celeste}}}\ (\bibinfo
  {publisher} {Bachelier et Huzard},\ \bibinfo {year} {1821})\BibitemShut
  {NoStop}%
\bibitem [{\citenamefont {{Capozziello}}\ and\ \citenamefont {{de
  Laurentis}}(2011)}]{capozziello}%
  \BibitemOpen
  \bibfield  {author} {\bibinfo {author} {\bibnamefont {{Capozziello}},
  \bibfnamefont {S.}}\ and\ \bibinfo {author} {\bibnamefont {{de Laurentis}},
  \bibfnamefont {M.}},\ }\href {\doibase 10.1016/j.physrep.2011.09.003}
  {\bibfield  {journal} {\bibinfo  {journal} {\physrep}\ }\textbf {\bibinfo
  {volume} {509}},\ \bibinfo {pages} {167} (\bibinfo {year} {2011})},\ \Eprint
  {http://arxiv.org/abs/arXiv:1108.6266} {arXiv:arXiv:1108.6266 [gr-qc]}
  \BibitemShut {NoStop}%
\bibitem [{\citenamefont {{Famaey}}\ and\ \citenamefont
  {{McGaugh}}(2012)}]{benoit}%
  \BibitemOpen
  \bibfield  {author} {\bibinfo {author} {\bibnamefont {{Famaey}},
  \bibfnamefont {B.}}\ and\ \bibinfo {author} {\bibnamefont {{McGaugh}},
  \bibfnamefont {S.~S.}},\ }\href {\doibase 10.12942/lrr-2012-10} {\bibfield
  {journal} {\bibinfo  {journal} {Living Reviews in Relativity}\ }\textbf
  {\bibinfo {volume} {15}},\ \bibinfo {pages} {10} (\bibinfo {year} {2012})},\
  \Eprint {http://arxiv.org/abs/arXiv:1112.3960} {arXiv:arXiv:1112.3960
  [astro-ph.CO]} \BibitemShut {NoStop}%
\bibitem [{\citenamefont {{Gentile}}, \citenamefont {{Famaey}},\ and\
  \citenamefont {{de Blok}}(2011)}]{gentile}%
  \BibitemOpen
  \bibfield  {author} {\bibinfo {author} {\bibnamefont {{Gentile}},
  \bibfnamefont {G.}}, \bibinfo {author} {\bibnamefont {{Famaey}},
  \bibfnamefont {B.}}, \ and\ \bibinfo {author} {\bibnamefont {{de Blok}},
  \bibfnamefont {W.~J.~G.}},\ }\href {\doibase 10.1051/0004-6361/201015283}
  {\bibfield  {journal} {\bibinfo  {journal} {\aap}\ }\textbf {\bibinfo
  {volume} {527}},\ \bibinfo {eid} {A76} (\bibinfo {year} {2011})},\ \Eprint
  {http://arxiv.org/abs/arXiv:1011.4148} {arXiv:arXiv:1011.4148 [astro-ph.CO]}
  \BibitemShut {NoStop}%
\bibitem [{\citenamefont {{Hernandez}}\ and\ \citenamefont
  {{Lee}}(2010)}]{hernandez10}%
  \BibitemOpen
  \bibfield  {author} {\bibinfo {author} {\bibnamefont {{Hernandez}},
  \bibfnamefont {X.}}\ and\ \bibinfo {author} {\bibnamefont {{Lee}},
  \bibfnamefont {W.~H.}},\ }\href {\doibase 10.1111/j.1745-3933.2010.00823.x}
  {\bibfield  {journal} {\bibinfo  {journal} {\mnras}\ }\textbf {\bibinfo
  {volume} {404}},\ \bibinfo {pages} {L6} (\bibinfo {year} {2010})},\ \Eprint
  {http://arxiv.org/abs/1002.0553} {arXiv:1002.0553 [astro-ph.CO]} \BibitemShut
  {NoStop}%
\bibitem [{\citenamefont {{Iocco}}, \citenamefont {{Pato}},\ and\ \citenamefont
  {{Bertone}}(2015)}]{iocco}%
  \BibitemOpen
  \bibfield  {author} {\bibinfo {author} {\bibnamefont {{Iocco}}, \bibfnamefont
  {F.}}, \bibinfo {author} {\bibnamefont {{Pato}}, \bibfnamefont {M.}}, \ and\
  \bibinfo {author} {\bibnamefont {{Bertone}}, \bibfnamefont {G.}},\ }\href
  {\doibase doi:10.1038/nphys3237} {\bibfield  {journal} {\bibinfo  {journal}
  {Nature Physics}\ } (\bibinfo {year} {2015}),\ doi:10.1038/nphys3237},\
  \bibinfo {note} {advanced online publication},\ \Eprint
  {http://arxiv.org/abs/doi:10.1038/nphys3237} {doi:10.1038/nphys3237}
  \BibitemShut {NoStop}%
\bibitem [{\citenamefont {{Mendoza}}\ \emph {et~al.}(2011)\citenamefont
  {{Mendoza}}, \citenamefont {{Hernandez}}, \citenamefont {{Hidalgo}},\ and\
  \citenamefont {{Bernal}}}]{mendoza}%
  \BibitemOpen
  \bibfield  {author} {\bibinfo {author} {\bibnamefont {{Mendoza}},
  \bibfnamefont {S.}}, \bibinfo {author} {\bibnamefont {{Hernandez}},
  \bibfnamefont {X.}}, \bibinfo {author} {\bibnamefont {{Hidalgo}},
  \bibfnamefont {J.~C.}}, \ and\ \bibinfo {author} {\bibnamefont {{Bernal}},
  \bibfnamefont {T.}},\ }\href {\doibase 10.1111/j.1365-2966.2010.17685.x}
  {\bibfield  {journal} {\bibinfo  {journal} {MNRAS}\ }\textbf {\bibinfo
  {volume} {411}},\ \bibinfo {pages} {226} (\bibinfo {year} {2011})},\ \Eprint
  {http://arxiv.org/abs/arXiv:1006.5037} {arXiv:arXiv:1006.5037 [astro-ph.GA]}
  \BibitemShut {NoStop}%
\bibitem [{\citenamefont {{Milgrom}}(1983)}]{milgrom}%
  \BibitemOpen
  \bibfield  {author} {\bibinfo {author} {\bibnamefont {{Milgrom}},
  \bibfnamefont {M.}},\ }\href {\doibase 10.1086/161131} {\bibfield  {journal}
  {\bibinfo  {journal} {\apj}\ }\textbf {\bibinfo {volume} {270}},\ \bibinfo
  {pages} {371} (\bibinfo {year} {1983})}\BibitemShut {NoStop}%
\bibitem [{\citenamefont {{Nojiri}}\ and\ \citenamefont
  {{Odintsov}}(2011)}]{odintsov}%
  \BibitemOpen
  \bibfield  {author} {\bibinfo {author} {\bibnamefont {{Nojiri}},
  \bibfnamefont {S.}}\ and\ \bibinfo {author} {\bibnamefont {{Odintsov}},
  \bibfnamefont {S.~D.}},\ }\href {\doibase 10.1016/j.physrep.2011.04.001}
  {\bibfield  {journal} {\bibinfo  {journal} {\physrep}\ }\textbf {\bibinfo
  {volume} {505}},\ \bibinfo {pages} {59} (\bibinfo {year} {2011})},\ \Eprint
  {http://arxiv.org/abs/arXiv:1011.0544} {arXiv:arXiv:1011.0544 [gr-qc]}
  \BibitemShut {NoStop}%
\bibitem [{\citenamefont {{S{\'a}nchez-Salcedo}}, \citenamefont
  {{Reyes-Iturbide}},\ and\ \citenamefont {{Hernandez}}(2006)}]{hernandez06}%
  \BibitemOpen
  \bibfield  {author} {\bibinfo {author} {\bibnamefont {{S{\'a}nchez-Salcedo}},
  \bibfnamefont {F.~J.}}, \bibinfo {author} {\bibnamefont {{Reyes-Iturbide}},
  \bibfnamefont {J.}}, \ and\ \bibinfo {author} {\bibnamefont {{Hernandez}},
  \bibfnamefont {X.}},\ }\href {\doibase 10.1111/j.1365-2966.2006.10602.x}
  {\bibfield  {journal} {\bibinfo  {journal} {\mnras}\ }\textbf {\bibinfo
  {volume} {370}},\ \bibinfo {pages} {1829} (\bibinfo {year} {2006})},\ \Eprint
  {http://arxiv.org/abs/astro-ph/0601490} {astro-ph/0601490} \BibitemShut
  {NoStop}%
\bibitem [{\citenamefont {{Springel}}\ \emph {et~al.}(2008)\citenamefont
  {{Springel}}, \citenamefont {{Wang}}, \citenamefont {{Vogelsberger}},
  \citenamefont {{Ludlow}}, \citenamefont {{Jenkins}}, \citenamefont {{Helmi}},
  \citenamefont {{Navarro}}, \citenamefont {{Frenk}},\ and\ \citenamefont
  {{White}}}]{dm}%
  \BibitemOpen
  \bibfield  {author} {\bibinfo {author} {\bibnamefont {{Springel}},
  \bibfnamefont {V.}}, \bibinfo {author} {\bibnamefont {{Wang}}, \bibfnamefont
  {J.}}, \bibinfo {author} {\bibnamefont {{Vogelsberger}}, \bibfnamefont {M.}},
  \bibinfo {author} {\bibnamefont {{Ludlow}}, \bibfnamefont {A.}}, \bibinfo
  {author} {\bibnamefont {{Jenkins}}, \bibfnamefont {A.}}, \bibinfo {author}
  {\bibnamefont {{Helmi}}, \bibfnamefont {A.}}, \bibinfo {author} {\bibnamefont
  {{Navarro}}, \bibfnamefont {J.~F.}}, \bibinfo {author} {\bibnamefont
  {{Frenk}}, \bibfnamefont {C.~S.}}, \ and\ \bibinfo {author} {\bibnamefont
  {{White}}, \bibfnamefont {S.~D.~M.}},\ }\href {\doibase
  10.1111/j.1365-2966.2008.14066.x} {\bibfield  {journal} {\bibinfo  {journal}
  {\mnras}\ }\textbf {\bibinfo {volume} {391}},\ \bibinfo {pages} {1685}
  (\bibinfo {year} {2008})},\ \Eprint {http://arxiv.org/abs/arXiv:0809.0898}
  {arXiv:arXiv:0809.0898} \BibitemShut {NoStop}%
\end{thebibliography}%

\end{document}